\def\kp  {$K^{\prime}$}
\def\kms{\ifmmode{\hbox{km~s}^{-1}}\else{km~s$^{-1}$}\fi}
\def\la{\mathrel{\hbox{\rlap{\hbox{\lower4pt\hbox{$\sim$}}}\raise1pt\hbox{$<$}}}}
\def\ga{\mathrel{\hbox{\rlap{\hbox{\lower4pt\hbox{$\sim$}}}\raise1pt\hbox{$>$}}}}
\documentstyle[aaspp4]{article} %single-spaced preprint style
\begin{document}
\title {\bf The Correlation Between Galaxy HI Linewidths and \kp\ Luminosities}

\author{Barry Rothberg$^1$, Will Saunders$^2$, R. Brent Tully$^1$, and Peter L.
Witchalls$^{2,3}$}

\affil {$^1$ Institute for Astronomy, University of Hawaii, 
2680 Woodlawn Drive, Honolulu, HI 96822 e-mail: tully@ifa.hawaii.edu}
\affil {$^2$ Institute for Astronomy, University of Edinburgh, Royal Observatory,
Blackford Hill, Edinburgh EH9 3HJ, Scotland, UK}
\affil {$^3$ Deceased}

\begin{abstract}

The relationship between galaxy luminosities and rotation rates is studied
with total luminosities in the \kp\ band.  Extinction problems are essentially
eliminated at this band centered at $2.1 \mu$m.  A template 
luminosity--linewidth relation is derived based on 65 galaxies drawn from
two magnitude-limited cluster samples.  The zero-point is determined using
4 galaxies with accurately known distances.  The calibration is applied
to give the distance to the Pisces Cluster (60~Mpc) at a redshift in the
CMB frame of 4771~\kms.  The resultant value of the Hubble Constant is
81~\kms Mpc$^{-1}$.  The largest sources of uncertainty arises from 
the small number of zero-point calibrators at this time at \kp\ and
present application to only one cluster.

\end{abstract}


\section{Luminosity -- Linewidth Correlations}

There is a strong correlation between the total luminosities of galaxies 
and the global rotation velocities (Tully \& Fisher 1977). \markcite{tfi}  
The correlation can
be applied to give distances to galaxies.  The 
rotation parameter gives an expectation luminosity and the apparent 
luminosity falls off from this value as the square of the distance.  
Aaronson, Huchra, \& Mould \markcite{aa} (1979) noted the
advantages of {\it infrared} luminosities: obscuration
is minimized and the light comes from old stars 
which have less stochastic variability than the young stars. There is a 
substantial correlation between color and luminosity for spirals so the 
detailed 
correlations between luminosity and the rotation rate parameter depend 
on wavelength (Tully, Mould, \& Aaronson \markcite{tm} 1982), steepening toward the 
infrared.

The original demonstration of the correlation used photographic or $B$ band
data while the Aaronson et al. collaboration pioneered work at aperture
$H$ band ($1.6 \mu$m).  In recent years, there has been a great deal of
work at intermediate bands; either in the red where the contrast against
the sky background is optimal (Pierce \& Tully 1988, \markcite{pt1} 1992; 
\markcite{pt2} Willick 1991; \markcite{wi}
Courteau 1996 \markcite{co}) or at $I$ band where obscuration effects are the most
diminished within the spectral domain accessible to CCD cameras
(Pierce \& Tully 1988, 1992; Han 1992; \markcite{ha} Mathewson, Ford, \& 
Buchhorn 1992; \markcite{mat}
Giovanelli et al 1997). \markcite{gio}

Since the seminal aperture photometry work by Aaronson et al., there has
been only limited work at longer wavelengths (eg, Peletier \& Willner
1991, \markcite{pe1} 1993) \markcite{pe2} although electronic imagers have 
become available with 
reasonably large formats.  The primary disadvantage at wavelengths 
greater than 1~micron is the strong and variable atmospheric emission.
All but the very centers of the highest surface brightness galaxies
emit below the flux levels of the sky.  Our experience is at \kp,
($2.1 \mu$m) a shift of the $K$ band to avoid thermal emission (Wainscoat 
\& Cowie 1992). \markcite{wa}  One observes several scalelengths
farther out in a galaxy at $R$ or $I$ than at \kp\ with a fixed integration
at the same telescope focus (Tully et al. 1996). \markcite{tv2}  On the other hand,
there is the considerable advantage that obscuration problems are almost
eliminated by \kp\ (25\% of the $I$ extinction according to Tully et al. 
1998). \markcite{tp2}  If
extinction can be largely negated then one can restrict to samples that 
are highly inclined which minimizes the corrections required for 
projection effects in the kinematic parameter.  Arguably the most 
dangerous 
potential systematic effects arise out of uncertain inclination and 
extinction corrections.  Hence, although the correlations at \kp\ are 
no better than at $I$ and are a lot more work to acquire, they do eliminate
some systematic problems.  The first order of business is to see if one
gets the same results across the spectrum from blue to infrared bands.

\section{\kp\ Photometry of the Separate Samples}

Our strategy parallels the optical band procedures described by
Tully \& Pierce (1999; hereafter TP). \markcite{tp}  Template luminosity -- HI 
profile linewidth 
correlations are produced for magnitude-limited cluster samples.
In the present case, data are available for two clusters.  We vary
the assumed differential distance modulus between clusters until an rms 
minimization is achieved with a linear regression slope to the common
data set.  The regression takes errors to be in linewidths, the
prescription described by TP to avoid the Malmquist (1920) bias
caused by a magnitude limit.
The absolute scale of the correlation is set by least squares fit of
the template relation to the luminosity--linewidth properties of a
small number of galaxies with accurate independent distances.

\subsection{The Ursa Major Cluster}

The Ursa Major Cluster is as close to home as the Virgo or Fornax 
clusters but
it is much more irregular, with no defined center, and a dynamical 
collapse
time not much less than the age of the universe.  The 79 known members
are predominantly HI-rich disk systems, similar to those found in low
density environments.  The sample was discussed by Tully et al. (1996)
and the data used in the current paper are presented in that earlier 
publication.  There are 38 galaxies with $I<13.4$ (after corrections), 
inclination 
$i\ge45^{\circ}$ and type S$a$ or later.  The same sample is used to
study optical band luminosity--linewidth correlations by TP and we use 
the same HI profile and inclination information.
The small corrections for extinction as a function of galaxy inclination
are those described by Tully et al.
(1998).  The tiny corrections for Galactic extinction are derived from
the selected reddening given by Schlegel, Finkbeiner, \& Davis (1998);
\markcite{scl}
we take $A_{K^{\prime}}^b=0.37E(B-V)$.
The relevant data are gathered together in Table~1.

\subsection{The Pisces Cluster}

The sample to be considered is a spatially restricted subset of the
material discussed by TP.  The Perseus-Pisces
filament is a prominent structure at roughly 5,000~\kms\ that includes
several Abell and other clusters.  The study of optical correlations
by TP drew galaxies from a $20^{\circ}$ length of the filament.
Here we consider 27 galaxies drawn from the restricted range
$0^h49^m < R.A. < 1^h32^m$.  There are several sub-condensations of
galaxies in this so-called Pisces Cluster but the structure is 
irregular and probably dynamically young overall.  Sakai, Giovanelli,
\& Wegner \markcite{sak} (1994) identify separate NGC~383 and NGC~507 groups within
this region.  TP failed to find any differential in distances
between separate components.

The \kp\ photometry to be used is published here for the first time 
although it was involved in the analysis by Tully et al. (1998).
The observations were made on 10 November, 1995, with the University of
Hawaii 2.2m telescope on Mauna Kea.  A 1024x1024 HgCdTe detector provided
a field of 193 arcsec with 0.19 arcsec pixels.  Five dithered 60 sec
exposures were combined.  The observing sequence would run
sky - object 1 - object 2 - sky - object3 - object4 - sky ...
Hence object observations were always either immediately preceded or
followed by sky observations so there could be proper subtraction of the
sky background.  Pixel to pixel variations are eliminated using a 
combination of dome and sky flats.
Dithered images were registered and median filtered to
eliminate defects.

Reductions were carried out with IRAF routines.  Once contaminating
sources were masked, elliptical isophotes were fit progressively with
radius, exponential disks were fit, and fluxes were extrapolated to
infinity to give total magnitudes.

The small reddening corrections follow the same prescriptions mentioned
in the previous sub-section.
The HI profile information and inclinations are again the same as those
used in the optical study of TP.  The relevant material is 
presented in Table~1.

\subsection{Zero-Point Calibrators}

A special set-up provides a very wide field for \kp\ observations.
The University of Hawaii 0.61m telescope is used to piggyback an 0.25m
telescope that feeds the 1024x1024 HgCdTe detector.  The set-up gives a
29 arcmin field with 1.69 arcsec pixels.  This field of view is greater
than the dimensions of all
galaxies in the northern sky with the exceptions of M31 and M33.

Observations of many of the largest galaxies were undertaken 
6-11 February, 1997.  Among the galaxies successfully observed, there are
four that have accurate independent distances.  Cepheid 
variables have been observed in NGC~3031 (Freedman et al. 1994: \markcite{fr}
$(m-M)_0=27.80$),
NGC~3198 (Kelson et al. 1999: \markcite{ke} $(m-M)_0=30.80$), NGC~3627 
(Saha et al. 1999; \markcite{sah} Gibson et al. 1999: \markcite{gib}
$(m-M)_0=30.06$), and NGC~4258 (Maoz et al. 1999: \markcite{mao}
$(m-M)_0=29.54$).  HI line profile 
information is available for each of these galaxies.  The 4 galaxies are 
among the expanded 
set of calibrators used in the TP optical band study.

The observing strategy with the small telescope is similar to that used
with the 2.2m telescope except the exposures extend over an hour per 
object so more frequent excursions are needed between object and sky.
We observe in the pattern
object - sky north - object - sky south - object - sky east - object - 
sky west (5 object, 4 sky).  Exposures were 120 seconds and the pattern
was repeated twice per object, resulting in a cumulative on-target
exposure of 20 minutes.

Reduction procedures are similar between the observations taken with 
the large and small telescopes except for a caveat about flat fielding.
Dome flats are not possible with the 0.25m telescope because the dome
is not sufficiently out of focus.  The observations with the small
telescope were obtained remotely and we were not successful with
twilight flats during the run in question.  Hence flats were exclusively
generated by median filtering all the off-object frames.

Adjustments for inclination and extinction within our galaxy follow the
same recipes previously mentioned.  Again, inclination and HI linewidth
information is taken from TP.  This material is gathered in Table~1.

\section{Tabulated Data}

The data is accumulated in Table~1 and provides information comparable 
to that in a table in TP.  The following information is provided
in each column. (1) Principal Galaxies Catalogue (PGC) number from the 
Lyon Extragalactic Database. (2) Alternative
names; by preference, NGC (N), UGC (U), Zwicky (Z).
(3) Morphological types (T:1,3,5,7,9=Sa,Sb,Sc,Sd,Sm). (4) Systemic
velocity in the rest frame of the cosmic microwave background.
(5) Galactic foreground extinction coefficient, $E(B-V)$.
(6) Axial ratio of minor axis to major axis, $q$. (7) Inclination, $i$.
(8) Total magnitudes, \kp$_T$.
(9) Total magnitudes adjusted for galactic extinction ($b$), 
inclination-dependent extinction ($i$), and $k$-correction ($k$), 
\kp$_T^{bik}$.
(10) Absolute magnitudes at the accepted distance modulus,
$M_{K^{\prime}}^{bik}$.
(11) HI linewidth, $W_{20}$. (12) Linewidth uncertainty.
(13) Logarithm of adjusted linewidth, ${\rm log} W_R^i$.
(14) References for HI linewidths.

HI linewidth references are given by a 3 figure code.  If the code is less
than 600 then the reference is provided by Huchtmeier \&  Richter
\markcite{hu} (1989)
for that code.  We have been maintaining a database that follows on from
Huchtmeier \&  Richter and the additional references of concern are given
here: 601 $=$ Begeman \markcite{be} (1989),
619 $=$ Magri \markcite{mag} (1990),
630 $=$ Haynes \& Giovanelli \markcite{hg1} (1991$a$), 631 $=$ Haynes \& 
Giovanelli \markcite{hg2} (1991$b$), 637 $=$ Roth et al. (1991),
\markcite{ro}
655 $=$ Schneider et al. \markcite{scn} (1992),
660 $=$ Broeils \markcite{br} (1992),
700 $=$ Wegner et al. \markcite{we} (1993),
701 $=$ Haynes et al. \markcite{hg3} (1997), 702 $=$ Haynes, private communication,
706 $=$ Giovanelli et al. (1997),
707 $=$ Tully \& Verheijen \markcite{tv} (1997).

\section{The \kp\ Luminosity--Linewidth Template Relation}

In the two panels of Figure~1, one sees apparent magnitudes plotted 
against a logarithmic linewidth parameter for the two separate clusters.
The linewidth parameter is derived from the 21~cm HI profile width and
approximates $2 V_{max}$, where $V_{max}$ is the rotation curve maximum
velocity (Tully \& Fouqu\'e 1985). \markcite{tfo}  Within type and inclination limits,
the Ursa Major sample is complete to $I_T^{bik}=13.4^m$ and the Pisces
sample is almost complete to $I_T^{bik}=13.8^m$ (where $I_T$ magnitudes 
are extrapolated to infinity and the superscripts indicate corrections
have been made for Galactic and internal extinction and the redshift 
effect).

In Figure~2, the Pisces sample has been shifted $2.55^m$ to achieve a 
best fit with the Ursa Major data.  The fitting procedure is described 
in TP.  Here is a brief review.  Least squares fits are made with 
errors taken in linewidths, the procedure that nulls the Malmquist bias
that arises with magnitude-limited samples.  (Naive application of the
regression with errors in magnitude to give distances would result in biased 
distances $\sim10\%$ low.)
A fit is made to the Ursa 
Major data alone, and the slope is force fit to the Pisces data to obtain a
preliminary offset.  Once this offset is applied, a fit is made to the
combined data sets, then the new fit is applied separately to the two
clusters to determine a residual difference.  The Pisces offset is 
corrected by the residual, the process is repeated, and convergence is
rapidly achieved.  The best fit is illustrated by the dashed line in 
Fig.~2.  The rms scatter of the 65 points is $\pm 0.44^m$.

\section{The Zero--Point Calibration}

Accurate distances are known for 4 galaxies with \kp\ apparent magnitude
and HI linewidth information: NGC~3031, NGC~3198, NGC~3627, and 
NGC~4258.  The intrinsic luminosity--linewidth relation for the four
systems is shown in Figure~3.  The dashed line has the slope derived from
the Ursa Major + Pisces template and a zero-point that minimizes the sum
of the deviations squared.  This line is described by the formula:
\begin{equation}
M_{K^{\prime}}^{bik} = -23.13(\pm0.12) - 8.78(\pm0.33) ({\rm log}W_R^i - 2.5)
\end{equation}
The $1\sigma$ error on the slope is derived from the 2 cluster fit.
The error on the zero-point is derived from the constraint from the 4
galaxies with Cepheid distances.
The rms scatter of the four points is $\pm 0.22^m$.

The fit of the template data to the absolute calibration is seen in 
Figure~4.  This fit requires a distance modulus for the Ursa Major
Cluster of 31.31 and a modulus for Pisces of 33.86.

\section{Distances to Additional Galaxies}

Ten additional galaxies were observed at \kp\ with the 0.25m aperture
wide field arrangement.  The absolute calibration of the \kp\
luminosity--linewidth correlation given by Eq.~(1) can be used to 
establish the distances of these galaxies.  Information about these 
systems is provided in Table~2.  The format of this table resembles
that of Table~1 with small differences.  The systemic velocities
are the $V_{GSR}$ values of the Third Reference Catalogue (de Vaucouleurs
et al. 1991) \markcite{de} rather than velocities in the frame of the microwave 
background 
and the absolute magnitudes are calculated from Eq.~(1).
There are two additional HI profile references: 613 $=$ Cayatte et al. 
\markcite{ca} (1990) and 638 $=$ Rupen \markcite{ru} (1991).

From the scatter in the luminosity--linewidth plots,
the rms uncertainty for an individual modulus is judged to be 
$\pm 0.4^m$.
The corrected linewidth for NGC~2841 is larger than the value for any
of the calibrator or template systems.  Hence application to this 
galaxy involves an extrapolation of the luminosity--linewidth 
correlation. 

\section{The Hubble Constant}

The constraints on the Hubble Constant provided by TP are more 
restrictive but this first determination with the \kp\ correlation
is worth noting.  The Ursa Major Cluster is strongly disturbed from
Hubble expansion by its proximity to the Virgo Cluster but the Pisces
Cluster is distant enough that its velocity may predominantly reflect 
the Hubble flow.  The velocity of this cluster in the microwave
background frame is 4771~\kms\ (Han \& Mould 1992) \markcite{hm} and the distance
from the \kp\ calibration is 59~Mpc, whence 
H$_0 = 81$~\kms Mpc$^{-1}$.

The statistical errors in this determination of H$_0$ are 
dominated by the small number of zero-point calibrators.  A reduced
$\chi^2$ goodness of fit is presented in Figure~5.  The 95\% 
probability limits on the Hubble Constant {\it from this uncertainty
alone} are $\pm 9$.

There is limited value in exploring all the potential uncertainties
in the current analysis because the same issues are addressed in TP,
but in that paper we use many more template galaxies, more 
calibrators, and more 
clusters extending to larger velocities.  The present \kp\ sample is
a subset of the optical sample and we use the same HI profile and
inclination information.  With so many factors held the same, we are
able to see if there are any significant systematics due only to the
photometry.  In particular, the \kp\ version has little sensitivity 
to extinction.  The scale difference is a remarkably small
$0.03^m$, a 1\% distance effect (the $1\sigma$ uncertainty with the
use of only 
4 calibrators is 5\%).  The analysis at $B,R,I$ bands by TP gives 
H$_0=77$
with essentially the entire difference a result of extending the 
analysis to include twelve clusters.

\section{In Memorium}

Our colleague Peter Witchalls carried out the photometric reductions
of the Pisces sample.  Sadly, he died before he saw the results of his
effort.

\clearpage

\begin{deluxetable}{llcrccrrrrrrcl}
\scriptsize
\tablewidth{6.9in}
\tablenum{1}
\tablecaption{Data for 4 Calibrators and 2 Clusters\label{b}}
\tablecolumns{14}
\tablehead{
\colhead{PGC}& %1
\colhead{Name}& %2
\colhead{Ty}& %3
\multicolumn{1}{c}{$V_{\rm 3K}$}& %4
\multicolumn{1}{c}{$E(B-V)$}& %5
\multicolumn{1}{c}{b/a}& %6
\multicolumn{1}{c}{Inc}& %7
\colhead{$K^{\prime}_{T}$}& %8
\colhead{$K^{\prime{\rm bik}}_{T}$}& %9
\colhead{$M_{K^\prime}^{\rm bik}$}& %10
\colhead{$W_{20}$}& %11
\colhead{e$_{20}$}& %12
\colhead{$\log W^i_R$}& %13
\multicolumn{1}{c}{H{\sc I} References} %14
}
\startdata
\multicolumn{14}{c}{Zero-Point Calibrators: 4 galaxies with independent
distances} \\ \tablevspace{4pt}
\tablevspace{-1pt}
\tableline
\tablevspace{13pt}
28630& N~3031& 2&   48& 0.080& 0.54& 59.& 3.55&  3.44& $-$24.36&  444&  7&
2.676& 80 102 185 296\\
30197& N~3198& 5&  880& 0.013& 0.42& 68.& 7.79&  7.71& $-$23.09&  321&  5&
2.484& 80 183 373 480 523 601\\
34695& N~3627& 3& 1066& 0.032& 0.61& 54.& 5.87&  5.80& $-$24.26&  381&  5&
2.626& 80 113 183 373 473 515\\
39600& N~4258& 4&  657& 0.016& 0.36& 72.& 5.23&  5.10& $-$24.44&  442&  5&
2.628& 183 373 387 442 473 480\\
\cutinhead{Ursa Major: 38 galaxies, distance modulus = 31.31}\\
34971& U~6399&   9&  1002& 0.015& 0.28& 78.& 11.09& 11.04& $-$20.27&  173&
20& 2.153&  373\\
35202& U~6446&   7&   826& 0.016& 0.62& 53.& 11.50& 11.47& $-$19.84&  156&
9& 2.193&  157 373 706\\
35616& N~3718&   1&  1174& 0.014& 0.42& 68.&  7.47&  7.36& $-$23.95&  481&
5& 2.678& 80 372 373 417 441 442\\
35676& N~3726&   5&  1079& 0.017& 0.62& 53.&  7.96&  7.91& $-$23.40&  294&
6& 2.504& 80 203 373 387 429 480\\
35711& N~3729&   2&  1249& 0.011& 0.66& 50.&  8.60&  8.56& $-$22.75&  279&
15& 2.496&  441\\
35999& N~3769&   3&   947& 0.023& 0.33& 75.&  9.10&  9.01& $-$22.30&  263&
8& 2.366&  387 417 631\\
36343& U~6667&   6&  1170& 0.017& 0.15& 90.& 10.81& 10.72& $-$20.59&  199&
11& 2.212&  373 387 706\\
36699& N~3877&   5&  1114& 0.023& 0.22& 84.&  7.75&  7.60& $-$23.71&  359&
7& 2.507&  373 387 429 660
706\\
36825& U~6773&   9&  1124& 0.017& 0.50& 62.& 11.23& 11.21& $-$20.10&  119&
7& 2.026&  630 655\\
36875& N~3893&   5&  1176& 0.021& 0.65& 51.&  7.84&  7.79& $-$23.52&  312&
5& 2.546& 80 203 373 387 417
660\\
37036& N~3917&   6&  1158& 0.022& 0.23& 83.&  9.08&  8.95& $-$22.36&  295&
6& 2.412&  201 373 387 660
702\\
37037& U~6816&   9&  1055& 0.014& 0.69& 47.& 11.91& 11.89& $-$19.42&  141&
7& 2.184&  373 515\\
37038& U~6818&   7&  1033& 0.022& 0.28& 78.& 11.70& 11.65& $-$19.66&  176&
12& 2.161&  373 387\\
37290& N~3949&   4&  1009& 0.021& 0.63& 52.&  8.43&  8.38& $-$22.93&  286&
6& 2.495&  373 387 442 706\\
37306& N~3953&   4&  1241& 0.030& 0.50& 62.&  7.03&  6.94& $-$24.37&  426&
5& 2.641&  183 373 387 416
660 706\\
37466& N~3972&   4&  1014& 0.014& 0.28& 78.&  9.39&  9.29& $-$22.02&  266&
11& 2.367&  373 387\\
37525& U~6917&   7&  1113& 0.027& 0.58& 57.& 10.30& 10.26& $-$21.05&  202&
8& 2.295&  373 387\\
37542& N~3985&   9&  1157& 0.026& 0.70& 53.& 10.19& 10.16& $-$21.15&  179&
10& 2.257&  377 417 619\\
37553& U~6923&   8&  1248& 0.028& 0.40& 69.& 11.04& 10.99& $-$20.32&  174&
6& 2.176&  373 387 436 515\\
37617& N~3992&   4&  1230& 0.029& 0.50& 62.&  7.23&  7.13& $-$24.18&  479&
5& 2.697& 80 183 203 373 387
436\\
37691& N~4013&   3&  1064& 0.017& 0.22& 84.&  7.68&  7.51& $-$23.80&  409&
7& 2.570&  373 473 701\\
37697& N~4010&   7&  1116& 0.025& 0.16& 90.&  9.22&  9.08& $-$22.23&  276&
7& 2.375&  373 387 706\\
37719& U~6973&   2&   947& 0.021& 0.39& 70.&  8.23&  8.13& $-$23.18&  346&
10& 2.514&  707\\
37735& U~6983&   6&  1263& 0.027& 0.64& 52.& 10.52& 10.48& $-$20.83&  197&
7& 2.310&  157 373 387 706\\
38068& N~4051&   4&   928& 0.013& 0.75& 50.&  7.86&  7.83& $-$23.48&  267&
8& 2.475& 80 116 373\\
38283& N~4085&   5&   945& 0.018& 0.26& 80.&  9.20&  9.08& $-$22.23&  304&
7& 2.430&  373 387 407 442
619 706\\
38302& N~4088&   4&   952& 0.020& 0.38& 71.&  7.46&  7.35& $-$23.96&  373&
5& 2.548&  373 387 407 442
619\\
38356& U~7089&   8&  1007& 0.015& 0.20& 87.& 11.11& 11.06& $-$20.25&  159&
7& 2.104&  373 459\\
38370& N~4100&   4&  1272& 0.023& 0.30& 77.&  8.02&  7.88& $-$23.43&  404&
9& 2.573&  373 387 442 706\\
38375& U~7094&   8&  1011& 0.013& 0.36& 72.& 11.58& 11.57& $-$19.74&  112&
20& 1.967&  655\\
38392& N~4102&   2&  1021& 0.020& 0.57& 57.&  7.86&  7.80& $-$23.51&  328&
11& 2.537&  373 387 637\\
38643& N~4138&   1&  1105& 0.014& 0.63& 52.&  8.19&  8.14& $-$23.17&  329&
10& 2.566&  140 241 619\\
38795& N~4157&   3&   963& 0.021& 0.19& 90.&  7.52&  7.33& $-$23.98&  425&
7& 2.586&  373 387 706\\
38988& N~4183&   6&  1158& 0.015& 0.16& 90.&  9.76&  9.63& $-$21.68&  256&
7& 2.338&  201 373 706\\
39237& N~4218&   9&   923& 0.016& 0.60& 55.& 10.83& 10.80& $-$20.51&  156&
8& 2.182&  158 293 347 619\\
39241& N~4217&   3&  1234& 0.017& 0.27& 79.&  7.61&  7.46& $-$23.85&  421&
13& 2.589&  158 373 387 512 706\\
39285& N~4220&   1&  1136& 0.018& 0.31& 76.&  8.36&  8.23& $-$23.08&  372&
15& 2.535&  619\\
40537& N~4389&   4&   925& 0.015& 0.68& 49.&  9.12&  9.09& $-$22.22&  192&
8& 2.317&  346 373 387 619\\
\cutinhead{ Pisces filament: 27 galaxies, distance modulus = 33.86}
 2865& U~501&  5& 4769& 0.061& 0.14& 90.& 10.31& 10.08& $-$23.78&  404&  8&
2.557& 452 543\\
 2899& U~509&  5& 4825& 0.061& 0.42& 68.& 12.41& 12.33& $-$21.53&  236& 10&
2.324& 452\\
 2928& U~511&  5& 4287& 0.060& 0.22& 85.& 12.25& 12.10& $-$21.76&  310&  9&
2.430& 384 452\\
2964& Z501-024& 5& 4686& 0.066& 0.22& 85.& 12.56& 12.43& $-$21.43&  257&
15& 2.336& 565\\
 3020& U~525&  3& 4621& 0.060& 0.59& 56.& 11.52& 11.46& $-$22.40&  242&  7&
2.386& 452 543\\
 3108& U~540&  3& 4661& 0.053& 0.57& 57.& 11.13& 11.06& $-$22.80&  289&  7&
2.470& 452 543\\
 3133& U~542&  5& 4205& 0.056& 0.20& 90.& 10.00&  9.81& $-$24.05&  397&  7&
2.549& 452 543\\
 3222& U~557&  3& 4192& 0.060& 0.42& 68.& 11.90& 11.80& $-$22.06&  295&  8&
2.437& 452 543\\
 3235& U~556&  3& 4318& 0.057& 0.20& 90.& 10.28& 10.09& $-$23.77&  414&  8&
2.569& 452 543\\
 3260& N~295&  3& 5163& 0.062& 0.40& 69.&  9.05&  8.91& $-$24.95&  475& 15&
2.663& 543\\
 3274& N~296&  5& 5340& 0.064& 0.28& 78.& 11.72& 11.61& $-$22.25&  257&  9&
2.343& 452 543\\
 3336& U~575&  4& 4346& 0.061& 0.16& 90.& 11.51& 11.33& $-$22.53&  323& 17&
2.449& 452 543\\
 \tablebreak
 3606& U~623&  1& 4529& 0.059& 0.41& 69.& 10.66& 10.54& $-$23.32&  395& 15&
2.576& 452 543\\
 3611& N~338&  2& 4479& 0.055& 0.35& 73.&  9.04&  8.88& $-$24.98&  564&  8&
2.734& 452 543\\
 3664& U~633&  3& 5272& 0.064& 0.25& 82.& 10.76& 10.58& $-$23.28&  422&  7&
2.581& 452 543\\
 3866& U~669&  5& 5556& 0.070& 0.22& 85.& 10.95& 10.81& $-$23.05&  281& 15&
2.380& 452 543\\
 4110& U~714&  5& 4345& 0.064& 0.72& 45.& 10.86& 10.81& $-$23.05&  258& 15&
2.487& 452\\
 4561& N~444&  5& 4544& 0.064& 0.26& 80.& 11.52& 11.39& $-$22.47&  293&  8&
2.407& 452 543\\
 4563& U~809&  5& 3920& 0.059& 0.18& 90.& 11.37& 11.19& $-$22.67&  346& 14&
2.483& 543 700\\
 4596& N~452&  2& 4670& 0.066& 0.29& 78.&  9.40&  9.22& $-$24.64&  528& 12&
2.693& 452 543\\
 4735& U~841&  4& 5287& 0.062& 0.22& 84.& 11.48& 11.33& $-$22.53&  312&  8&
2.433& 452 543\\
 5035& N~494&  2& 5168& 0.061& 0.35& 73.&  9.35&  9.19& $-$24.67&  527& 15&
2.701& 452\\
 5061& N~496&  4& 5722& 0.072& 0.55& 59.& 10.53& 10.44& $-$23.42&  340& 20&
2.539& 452\\
 5284& U~987&  1& 4373& 0.060& 0.32& 75.&  9.72&  9.57& $-$24.29&  416&  8&
2.586& 452 543\\
 5344& N~536&  3& 4911& 0.052& 0.33& 74.&  8.80&  8.63& $-$25.23&  549& 10&
2.719& 543 700\\
 5440& U~1033& 5& 3757& 0.051& 0.18& 90.& 10.75& 10.56& $-$23.30&  367&  7&
2.512& 452 543\\
 5702& N~582&  3& 4070& 0.053& 0.24& 82.&  9.62&  9.43& $-$24.43&  474&  7&
2.638& 452 543\\
\enddata
\end{deluxetable}

\begin{deluxetable}{lccccccccccccl}
\scriptsize
%\tablewidth{7.67in}
\tablewidth{6.9in}
%Table 2. Distance Derivations for 10 Galaxies
% PGC   Name    Modulus type E(B-V) b/a inc   K'_T K'_T^bik
%W_20  log W_R^i HI Ref.
%                                                                M_K'^bik
%err_20
\tablenum{2}
\tablecaption{Distance Derivations for 10 Galaxies\label{t}}
\tablecolumns{14}
\tablehead{
\colhead{PGC}& %1
\colhead{Name}& %2
\colhead{Modulus}& %3
\colhead{Ty}& %4
%\colhead{V$_{\rm GSR}$}& %6
\colhead{$E(B-V)$}& %5
\colhead{b/a}& %6
\colhead{Inc}& %7
\colhead{$K^\prime_T$}& %8
\colhead{$K^{\prime\rm bik}_T$}& %9
\colhead{$M_{K^\prime}^{\rm bik}$}& %10
\colhead{$W_{20}$}& %11
\colhead{e$_{20}$}& %12
\colhead{$\log W^i_R$}& %13
\multicolumn{1}{c}{H{\sc I} References}\nl %14
%\cline{16}
}
\startdata
24930& N~2683&  30.09&  3&    0.033&  0.25&   81.&
        6.16&   5.99&   $-$24.09&       440& 5&      2.609&  183     373
        377     473     660\\
26512& N~2841&  31.85&  3&    0.015&  0.45&   66.&
        6.19&   6.07&   $-$25.77&       614& 4&      2.801&  183
        373     442     473     522\\
33550& N~3521&  30.31&  4&    0.058&  0.50&   62.&
        5.64&   5.53&   $-$24.77&       468& 5&      2.687&  183     373
        393     473     515\\
39028& N~4192&  31.08&  2&    0.035&  0.21&   86.&
        6.94&   6.74&   $-$24.33&       470& 5&      2.636&  151     373
        375     509     613\\
39422& N~4244&  28.67&  6&    0.021&  0.20&   90.&
        7.69&   7.59&   $-$21.07&       221& 7&      2.265&  373
        442\\
42002& N~4559&  29.03&  6&    0.018&  0.40&   69.&
        7.12&   7.05&   $-$21.97&       256& 5&      2.368&  201     373
        442     660\\
42038& N~4565&  30.42&   3&   0.015&  0.22&   85.&
        5.79&   5.59&   $-$24.82&       528& 5&      2.692&  183     373
        473     638\\
42637& N~4631&  28.79&  7&    0.017&  0.22&   85.&
        6.18&   6.04&   $-$22.74&       322& 6&      2.455&  183     222
        373\\
45948& N~5033&  31.07&  5&    0.012&  0.47&   64.&
        6.62&   6.52&   $-$24.54&       450& 4&      2.660&  183
        390     515     523     562\\
46153& N~5055&  29.54&  4&    0.017&  0.52&   61.&
        5.44&   5.36&   $-$24.17&       400& 5&      2.618&  183
        373     374     473     480\\
\enddata
\end{deluxetable}

\clearpage

\begin{figure}
\vspace{6.5cm}
\includegraphics{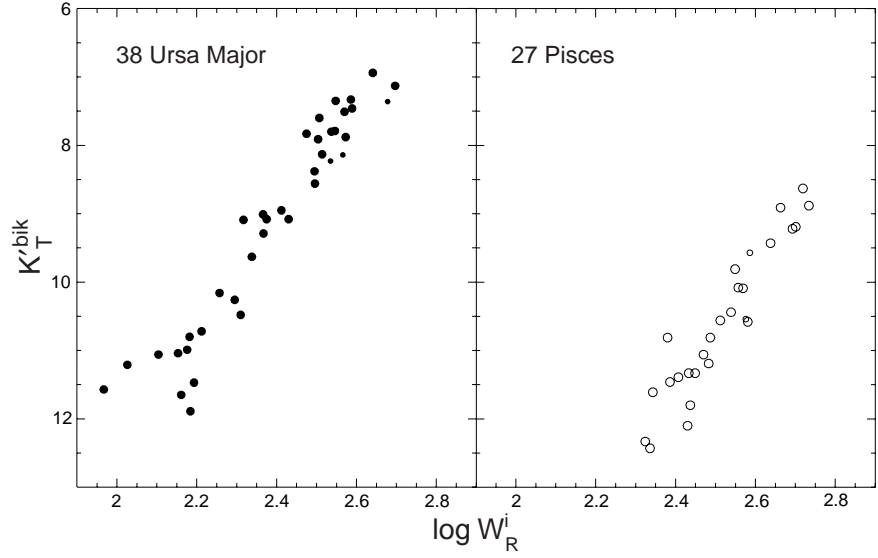}
\caption{
\kp-band apparent magnitude--HI profile linewidth plots for the two clusters 
that contribute to the template luminosity--linewidth correlations. 
Magnitudes are adjusted for internal and Galactic absorption and small
redshift corrections.
Large symbols: types S$ab$ and later.  Small symbols: type S$a$.
The Ursa Major sample is complete to $I_T^{bik}=13.4$.
The Pisces sample is nearly complete to
$I_T^{bik}=13.8$.  Galaxies fainter than these limits are excluded.
}\label{1}
\end{figure}

%\clearpage

\begin{figure}
\vspace{8.5cm}
\includegraphics{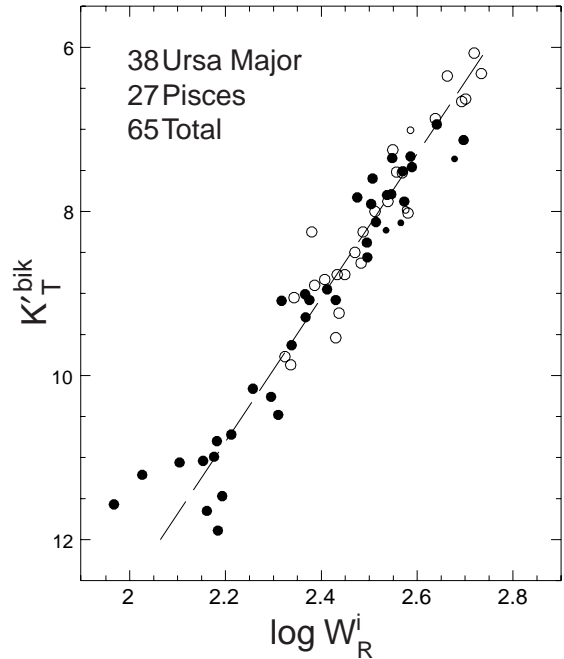}
\caption{
\kp-band apparent magnitude--linewidth relation for 2 clusters with the
Pisces sample translated $2.55^m$ to match 
the Ursa Major relation.  Symbols as in Fig.~1.  The straight line is a
least squares fit to the ensemble with errors in linewidths after the
iterations described in the text. 
}\label{2}
\end{figure}

%\clearpage

\begin{figure}
\vspace{7.5cm}
\includegraphics{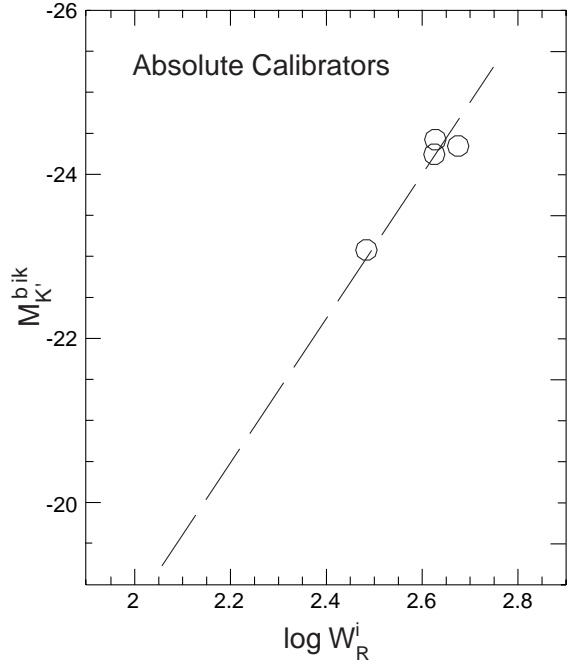}
\caption{
\kp\ absolute magnitude--linewidth relation for 4 galaxies with 
independently determined distances from
application of the cepheid period--luminosity relation.
The straight line is the least squares best fit slope shown in Fig.~2
shifted to fit the calibrators.
}\label{3}
\end{figure}

%\clearpage

\begin{figure}
\vspace{7.5cm}
\includegraphics{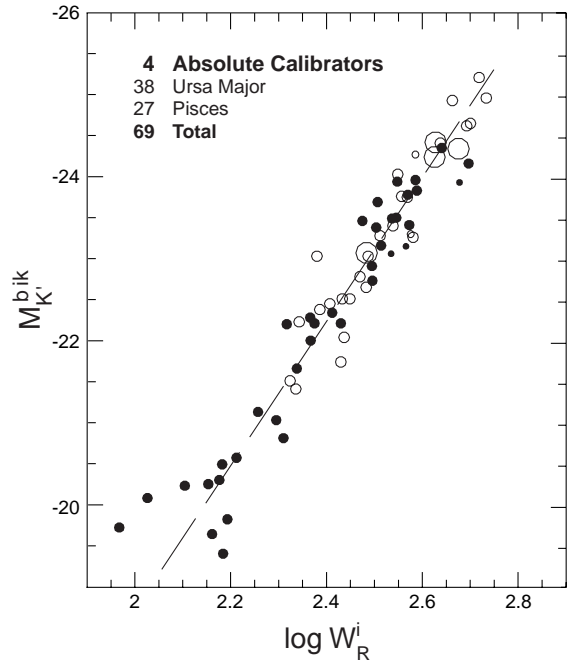}
\caption{
\kp\ absolute magnitude--linewidth relation with the cluster template
galaxies translated to overlay on the zero-point calibrator galaxies
($(m-M)_{UMa}=31.31$; $(m-M)_{Pisc}=33.86$).
Symbols and straight line fits as in previous plots.  
}\label{4}
\end{figure}

%\clearpage

\begin{figure}
\vspace{7cm}
\includegraphics{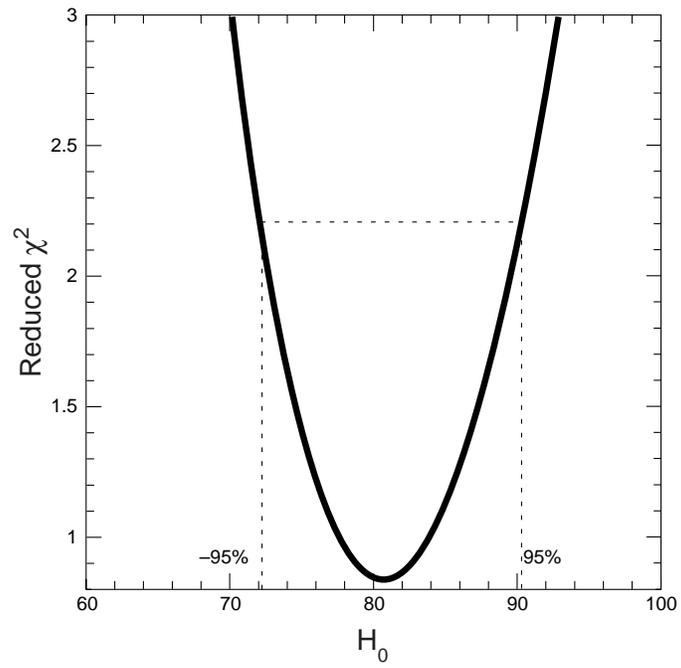}
\caption{
Reduced $\chi^2$ goodness of fit dependency on the choice of H$_0$ 
associated with the constraints provided by the 4 calibrators.
}\label{5}
\end{figure}

\end{document}